\documentclass[floatfix,
superscriptaddress,
onecolumn,
preprint,
aps,
pra]{revtex4-2}
\usepackage{amsmath, amsthm, amsfonts}
\usepackage{lipsum}
\usepackage{graphicx}
\usepackage{bm}
\usepackage{braket} 
\usepackage{changepage}
\usepackage{amsmath}
\usepackage{amssymb}
\usepackage[dvipsnames]{xcolor}

\DeclareFontFamily{OMS}{oasy}{\skewchar\font48 }
\DeclareFontShape{OMS}{oasy}{m}{n}{%
         <-5.5> oasy5     <5.5-6.5> oasy6
      <6.5-7.5> oasy7     <7.5-8.5> oasy8
      <8.5-9.5> oasy9     <9.5->  oasy10
      }{}
\DeclareFontShape{OMS}{oasy}{b}{n}{%
       <-6> oabsy5
      <6-8> oabsy7
      <8->  oabsy10
      }{}
\DeclareSymbolFont{oasy}{OMS}{oasy}{m}{n}
\SetSymbolFont{oasy}{bold}{OMS}{oasy}{b}{n}

\DeclareMathSymbol{\smallleftarrow}     {\mathrel}{oasy}{"20}
\DeclareMathSymbol{\smallrightarrow}    {\mathrel}{oasy}{"21}
\DeclareMathSymbol{\smallleftrightarrow}{\mathrel}{oasy}{"24}

\usepackage{changebar}

\begin{document}

\title{Upper bound to optical trapping stiffness through the multipolar control of optical beams}

\author{Enrique Ayllón-García}
\altaffiliation[]{enrique.ayllon@ehu.eus}
\affiliation{Centro de Física de Materiales (CFM), CSIC-EHU, Paseo Manuel de Lardizabal 5, 20018 Donostia-San Sebastián, Spain}
\affiliation{University of the Basque Country (EHU), Faculty of Science and Technology department of Electricity and Electronics}

\author{Iker Gómez-Viloria}
\affiliation{University of the Basque Country (EHU)}

\author{Quimey Pears Stefano}
\affiliation{Centro de Física de Materiales (CFM), CSIC-EHU, Paseo Manuel de Lardizabal 5, 20018 Donostia-San Sebastián, Spain}

\author{Jason T. Francis}
\affiliation{Centro de Física de Materiales (CFM), CSIC-EHU, Paseo Manuel de Lardizabal 5, 20018 Donostia-San Sebastián, Spain}
\affiliation{Donostia International Physics Center, Paseo Manuel de Lardizabal 4, 20018 Donostia-San Sebastián, Spain}

\author{Ángel Cifuentes}
\affiliation{LORTEK, Basque Research and Technology Alliance (BRTA), Arronamendia kalea 5A, 20240, Ordizia, Spain}

\author{Ásier Mongelos-Martinez}
\affiliation{Centro de Física de Materiales (CFM), CSIC-EHU, Paseo Manuel de Lardizabal 5, 20018 Donostia-San Sebastián, Spain}
\affiliation{University of the Basque Country (EHU), Faculty of Science and Technology department of Electricity and Electronics}

\author{Shah Jee Rahman}
\affiliation{Centro de Física de Materiales (CFM), CSIC-EHU, Paseo Manuel de Lardizabal 5, 20018 Donostia-San Sebastián, Spain}
\affiliation{University of the Basque Country (EHU), Faculty of Science and Technology department of Electricity and Electronics}
\affiliation{Donostia International Physics Center, Paseo Manuel de Lardizabal 4, 20018 Donostia-San Sebastián, Spain}

\author{Miguel Varga}
\affiliation{Centro de Física de Materiales (CFM), CSIC-EHU, Paseo Manuel de Lardizabal 5, 20018 Donostia-San Sebastián, Spain}

\author{Gabriel Molina-Terriza}
\altaffiliation[]{}
\affiliation{Centro de Física de Materiales (CFM), CSIC-EHU, Paseo Manuel de Lardizabal 5, 20018 Donostia-San Sebastián, Spain}
\affiliation{Donostia International Physics Center, Paseo Manuel de Lardizabal 4, 20018 Donostia-San Sebastián, Spain}
\affiliation{IKERBASQUE, Basque Foundation for Science, Maria Diaz de Haro 3, 48013 Bilbao, Spain}

\date{\today}

\begin{abstract}
{ \bf 

Optical tweezers enable the manipulation of microscopic objects using light, yet the fundamental limits to the optical forces that can be exerted on matter remain unknown. Here we derive a general upper bound to the maximum optical force that can be applied to a particle, based on an expansion of electromagnetic fields into well-defined  helicity multipolar modes. This method finds the optimal force for any kind of fields external to the particle, including evanescent fields. We apply the method to homogeneous spherical particles in a stable trap and identify the field distributions that saturate this bound for the trapping stiffness. We further provide experimentally accessible strategies to approach these optimal conditions, including configurations using counterpropagating and single-beam traps. Experiments demonstrate a threefold enhancement of trapping forces relative to conventional designs, while theoretical predictions indicate that order-of-magnitude improvements are achievable for larger particles and high angular momentum beams. Our results establish fundamental design principles for maximizing optical forces and define the ultimate limits of optical manipulation. }
\end{abstract}

\maketitle

\section{\label{sec:Intro}Main}

Since Arthur Ashkin’s first demonstration of optical trapping with a single beam in 1970 \cite{ashkin1986observation}, this technique has become an essential tool for particle manipulation, providing precise control over objects ranging from nanometers to micrometers in size \cite{pesce2020optical,volpe2023roadmap}. Optical trapping has been extended to the manipulation of atomic systems \cite{ashkin1979cooling,kaufman2021quantum}, biological samples \cite{ashok2012optical}, and particles in a variety of surrounding media \cite{summers2008trapping,stoellner2025using}. More recently, significant advances in optical levitation have enabled the control of particles in vacuum \cite{gonzalez2021levitodynamics}, reaching quantum regimes for mesoscopic objects isolated from environmental perturbations \cite{delic2020cooling,tebbenjohanns2021quantum}. A common challenge across these diverse platforms is that stable trapping typically requires high optical powers \cite{ashkin1986observation,gieseler2014dynamics}. Even weak absorption can therefore lead to substantial heating, which is particularly problematic for biological samples \cite{peterman2003laser,blazquez2019optical} or in levitated systems where heat dissipation is limited \cite{tebbenjohanns2021quantum}. Consequently, substantial effort has been devoted to improving the efficiency of optical trapping schemes.

Optical forces in trapping experiments are commonly quantified through the trapping stiffness ($\kappa$), an experimentally accessible parameter \cite{pesce2020optical,volpe2023roadmap} which characterizes how strongly a particle is restored to its equilibrium position. Increasing this stiffness is central to improving trapping efficiency and reducing the optical power required for stable confinement. Considerable effort has therefore been devoted to enhancing trapping stiffness using adaptive optics, beam shaping, and iterative optimization algorithms \cite{taylor2015enhanced,taylor2017optimizing,butaite2024photon}. However, despite these advances, no general method exists to determine whether such optimization strategies achieve the maximum possible trapping stiffness for a given optical power, except in limiting cases such as particles much smaller than the wavelength. As a result, it remains unclear whether current optical trapping schemes operate close to fundamental limits or whether substantial improvements remain possible.

A striking example of this uncertainty arises in the use of structured beams carrying orbital angular momentum \cite{molina2007twisted}. Experiments have shown that for particles comparable to or larger than the wavelength, such beams can generate stronger trapping forces than conventional diffraction-limited Gaussian beams \cite{o2001axial,gomez2024axis}. Remarkably, these particles may be attracted toward regions of reduced intensity, contrary to expectations based on simple intensity-gradient arguments \cite{gomez2024axis}. Although these observations demonstrate the potential of structured fields to enhance optical forces, the mechanisms responsible for this behavior remain incompletely understood, and it is unknown whether such configurations approach the ultimate performance limits. Numerical optimization methods may be advantageous for certain experimental conditions, but they provide little insight on the physical mechanisms behind the optimal optical beams, and may even miss some solutions outside of the optimization parameters. Thus, finding the fundamental bounds to optical trapping will considerably advance our understanding of optical forces and light matter interaction.

\begin{figure}[p]
\centering

\vspace*{-1.5cm}

\includegraphics[width=\textwidth]{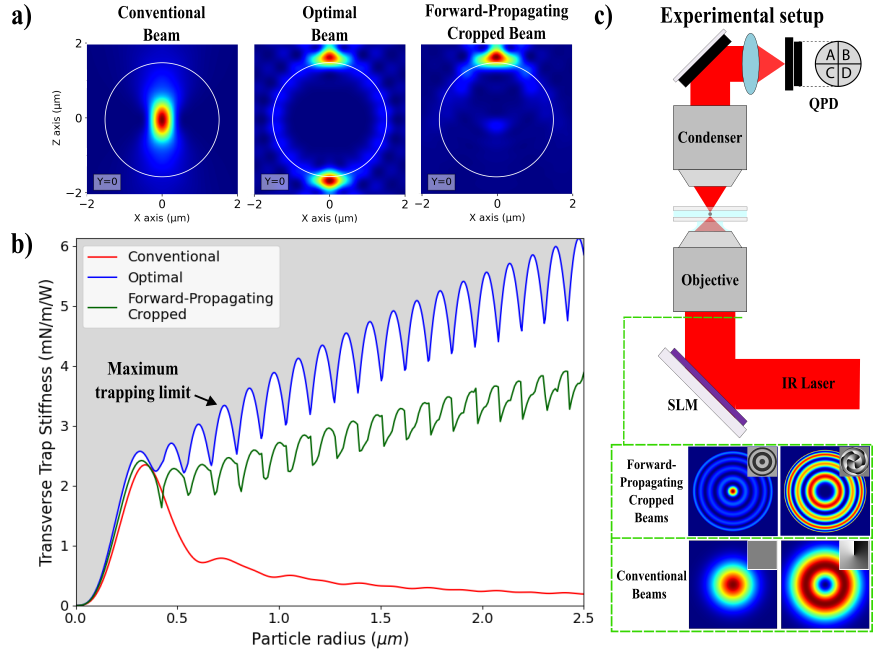}

\caption{Optimal beams for enhanced trap stiffness. (a) Comparison of the calculated beam intensities at focus in the Y=0 plane (propagation in the Z direction) for the three kind of beams considered in this work: a diffraction-limited conventional Gaussian beam used in optical trapping, the light field that would achieve the optimal trapping for the particle considered and an approximation using light propagating in the forward direction, i.e. forward-propagating cropped beams (FPC). The white circle shows a $3 \mu m$ $SiO_2$ particle used for this particular trapping example. (b) Transversal trap stiffness versus particle radius for $SiO_2$ particles in water illuminated with the three different beams studied in this work. Optimal beams set an upper bound to the achievable trap stiffness, such that the gray region indicates the range of trapping stiffness unaccessible to any kind of external beam. We are considering monochromatic light beams with a wavelength of 976 $nm$ and a well-defined helicity value of $p=1$ (see text for details). (c) Sketch of the experimental set-up used in the experiment to test the enhancement of the trapping stiffness. Our holographic optical tweezers setup contains a Spatial Light Modulator (SLM) for precise beam control and creating the beams needed to enhance the trapping stiffness. Inset: Calculated intensity distributions of the conventional and FPC beams at the objective's back focal aperture.}

\label{fig1}
\end{figure}

Here we resolve these questions by determining the optimal optical trapping conditions allowed by electromagnetic scattering. We derive exact upper bounds on the achievable trapping stiffness for a given object, based solely on its scattering properties, and identify the incident field distributions that attain these limits (Fig. 1). Our analysis reveals a hierarchy of trapping regimes that depends on particle size and shows that, for particles larger than the wavelength, optimal trapping requires structured fields with specific angular momentum content. The limits that we found apply to any optical trap with fields external to the particle, which include trapping schemes with evanescent fields. The only situations outside our framework are those where the particle interacts with the emitters inside the particle. These results establish the fundamental limits of optical trapping and provide a unified framework for understanding previously unexplained phenomena in structured-light trapping.

The optimal field configurations predicted by our theory generally involve illumination from multiple directions, which can be implemented experimentally using counter-propagating beams. To bridge the gap with conventional single-beam optical tweezers, we introduce a practical forward-propagating approximation that captures the dominant features of the optimal fields. These configurations provide experimentally accessible routes to approach the theoretical limits and enable substantial improvements in trapping stiffness over standard Gaussian-beam implementations.

Figure 1 illustrates the conceptual framework of our approach. Panel (a) compares conventional diffraction-limited beams with the optimal structured fields predicted by our theory, while panel (b) shows the   hierarchy of accessible trapping stiffness values as a function of particle size and how the optimal beams saturate the trapping stiffness bound. It also compares the calculated trapping stiffness for diffraction limited Gaussian beams and the Forward-Propagating Cropped (FPC) beam suitable for optimizing single beam optical traps. The experimental implementation used to approach these limits is also indicated schematically (panel (c)). We have found that our implementation with FPCs can attain trapping stiffness that triples the conventional diffraction limited Laguerre-Gaussian modes. Our experimental values are within 30\% of the theoretically calculated ones, which is compatible with aberrations in the system and inhomogeneous imperfections of the microscope objective. Importantly, stiffness increase was achieved without the use of iterative optimization methods to compensate for aberrations or experimental imperfections.

Our method is based on the expansion of optical forces using a well-defined helicity multipolar basis \cite{rose1955multipole,tischler2012role,gomez2025optical}, $\mathbf{A}_{j,m_z}^{p\,(1)}(\mathbf{r})$ and $\mathbf{A}_{j,m_z}^{p\,(3)}(\mathbf{r})$, for incident and scattered fields, respectively. The three indexes $p$, $j$ and $m_z$ correspond to the helicity ($p$), the multipolar order ($j$), which is the eigenmode of the squared total angular momentum $\textbf{J}^2=J_x^2+J_y^2+J_z^2$, and the total angular momentum ($m_z$), which is the eigenmode of the $z$ component $J_z$. In this framework, the incident field is decomposed in multipolar modes where each multipole is weighted by the coefficients known as Beam Shape Coefficients (BSC). Given that the helicity multipolar basis is complete for any electromagnetic fields, the only restriction for the incident fields is that it has a finite value inside the particle, i.e. it also includes external non-propagating fields, such as evanescent fields \cite{fernandez2016transverse}, only excluding emitters inside the particle. The use of well-defined helicity multipoles has two main advantages: 1) They have immediate applications to practical experimental situations when using focused circularly polarized beams \cite{tischler2014experimental}, as is the case of our experiments, and 2) the analytical expressions are very simple. For example, the expressions for the  BSCs of the paraxial and focused beams can be calculated  efficiently \cite{molina2008determination,zambrana2012excitationsinglemult}. Then, the incident and scattered fields, written in terms of the helicity multipoles are \cite{gomez2025optical}:
\begin{equation}
    \mathbf{E}_{in}= E_0\sum_{j=1}^\infty \sum_{m_z=-j}^j \, D_jC_{j,m_z,p} \mathbf{A}_{j,m_z}^{p\,(1)}
    \label{E_in_helicity}
\end{equation}
where $D_j=i^j\sqrt{2j+1}$ are the standard multipolar coefficients of plane waves and $C_{j,m_z,p}$ the BSCs \cite{molina2008determination,gomez2025optical}. In the following, we will set the BSCs in an ordered list in the form of a column vector: $\mathbf{C}=(C_{1,1,1},C_{2,1,1},...)^T$. On the other hand, the scattered  field, can be expressed as 
\begin{equation}
    \mathbf{E}_{sc}= E_0\sum_{j=1}^\infty \sum_{m_z=-j}^j \, D_jB_{j,m_z,p} \mathbf{A}_{j,m_z}^{p\,(3)}
    \label{E_sc_helicity}
\end{equation}
where, the coefficients $\mathbf{B}=(B_{1,1,1},B_{2,1,1},...)^T$, depend on the BSCs through the scattering matrix $\mathbf{\hat{S}}$: $\mathbf{B}=\mathbf{\hat{S}} \mathbf{C}$. 

This basis was previously used in Ref. \cite{gomez2025optical} to derive analytical expressions for the optical forces exerted by cylindrically symmetric focused beams with well-defined helicity impinging on spherical particles. With this formalism, the forces along a certain direction, say $\mathbf{u}$, acting on a particle displaced by a distance $\mathbf{d}$ with respect to the center of a beam described by a BSC vector $\mathbf{C}$ can be expressed as $F_\mathbf{u} = \mathbf{C}^\dagger \mathbf{\hat{F}}_\mathbf{u}(\mathbf{d})\mathbf{C}$. The force matrix $\mathbf{\hat{F}}_\mathbf{u}(\mathbf{d})$ has some interesting properties. It is a Hermitian matrix, because the forces acting on the particle are real, and it can be computed from the force matrix of the particle centered on the trapping beam: $\mathbf{\hat{F}}_\mathbf{u}(\mathbf{d})= \mathbf{\hat{\mathcal{T}}}^\dagger(\mathbf{d}) \mathbf{\hat{F}}_\mathbf{u}(\mathbf{0}) \mathbf{\hat{\mathcal{T}}}(\mathbf{d})$. Finally, it depends on the properties of the particle through the scattering matrix $\mathbf{\hat{S}}$. The translation matrices for the multipolar beams of well defined helicity are particularly simple \cite{tung1985group,zambrana2021excitationwhispering,gomez2025optical} and have an analytical expression for small displacements so for $\mathbf{d}=\epsilon \mathbf{v}$, with $\mathbf{v}$ an unitary vector, and $\epsilon$ small, $\mathbf{\hat{\mathcal{T}}}(\mathbf{d})=\mathbf{\hat{I}} + i \mathbf{\hat{P}}_\mathbf{v} \epsilon$, where $\mathbf{\hat{P}}_\mathbf{v}$ is the canonical momentum \cite{bliokh2015transverse,ghosh2024canonical,tung1985group}. With this we arrive at the expression of forces on arbitrary particles for small displacements:
\begin{equation}
F_{\textbf{u}}= \mathbf{C}^\dagger \mathbf{\hat{F}}_\mathbf{u}(\mathbf{0}) \mathbf{C} -  \epsilon \mathbf{C}^\dagger \mathbf{\hat{K}} \mathbf{C},
\label{Force_smalld}
\end{equation}
where $\mathbf{\hat{K}} = i [\mathbf{\hat{P}}_\mathbf{v}, \mathbf{\hat{F}}_\mathbf{u}(\mathbf{0}) ]$, with $[,]$ being the matrix commutation operation, is also a Hermitian matrix. In the case that the origin represents the equilibrium position, then $\mathbf{C}^\dagger \mathbf{\hat{F}}_\mathbf{u}(\mathbf{0}) \mathbf{C}=0$. Given that the trapping stiffness ($\kappa$) is by definition proportional to the displacement $\Delta x$ ($F=- \kappa\, \Delta x$), the matrix $\mathbf{\hat{K}}$ contains all the information about the trapping stiffness for arbitrary beams, i.e. $\kappa = \mathbf{C}^\dagger \mathbf{\hat{K}} \mathbf{C}$. The main result of this work relies on the fact that the eigenvalues of the matrix $\mathbf{\hat{K}}$ are real, being a Hermitian matrix, and that the maximum trapping stiffness is bounded by the highest eigenvalue, $\kappa_{OPT}$, which is \textbf{deterministically} achieved by the corresponding eigenstate $\mathbf{C}_{OPT}$. Importantly, given the scattering matrix of the object $\mathbf{\hat{S}}$, the trapping stiffness matrix, $\mathbf{\hat{K}}$, is completely determined. A few remarks are needed, before giving particular examples: It is also possible to consider out of equilibrium situations and maximize the pure force $\mathbf{C}^\dagger \mathbf{\hat{F}}_\mathbf{u}(\mathbf{0}) \mathbf{C}$, but in this work we are more interested in the trapping stiffness as it is the most common experimental parameter. Also, as explicitly written in Eq. (\ref{Force_smalld}), the force and the trapping stiffness matrix $\mathbf{\hat{K}}$ may be different for each of the force components $\mathbf{u}$ and the direction of the displacement $\mathbf{v}$. In general, optimizing the force along a given direction $\mathbf{u}$ will not be optimal for a different direction $\mathbf{u}^\prime$, an effect that has already been observed in levitation experiments \cite{rahman2025controlling}.

While this method is completely general, in this work we will restrict ourselves for simplicity to a very common experimental situation: spherical particles trapped using cylindrically symmetric beams and well defined helicity. Considering spherical particles simplifies the calculations as we can use the analytical expressions for the scattering of the particles from the Generalized Lorentz-Mie Theory \cite{gouesbet2011generalized}. For the case of spherical particles: $S_{j,m,p}^{k,n,q}=\alpha_j \delta_{j,k}\delta_{m,n}\delta_{p,q}+\beta_j \delta_{j,k}\delta_{m,n}\delta_{p,-q}$ where $\alpha_j$ and $\beta_j$ are the same- and cross-helicity scattering coefficients, respectively, and depend on the standard Mie coefficients, $a_j$ and $b_j$ \cite{bohren2008absorption}, as \cite{gomez2025optical}: \begin{equation}\label{alphabeta}
    \alpha_j=-\frac{a_j+b_j}{2},\;\;\beta_j=\frac{a_j-b_j}{2}
\end{equation}
The restriction on the incident fields means that we can focus on a subset of the multipolar fields, i.e. $\mathbf{C}=\mathbf{C}_{m_z,p}=(C_{1,m_z,p},C_{2,m_z,p},...,C_{j,m_z,p},...)^T$. These beams can be easily implemented in an experimental set-up, such as the one of Fig. 1 c), using focused beams with well defined angular momentum and circular polarization. Now, the maximal trapping stiffness (blue curve in Fig. 1 b) )  and the optimal beam (central figures of Fig. 1 a) ) have an operational meaning, given directly by the maximum eigenvalue and its eigenvector of the matrix $\mathbf{\hat{K}}$ obtained for $\mathrm{SiO_2}$ spheres suspended in water (index contrast of approximately 1.1). In addition, the experimental realization of the optimal beams can be obtained from the inverse method used to retrieve the focused BSC coefficients from incident optical beams \cite{molina2008determination,zambrana2012excitationsinglemult}.

\section{Optimal beam determination for $SiO_2$ particles}\label{Sec2}
\begin{figure}[p]
    \centering
    \vspace*{-1.5cm}
    \includegraphics[width=0.67\textwidth]{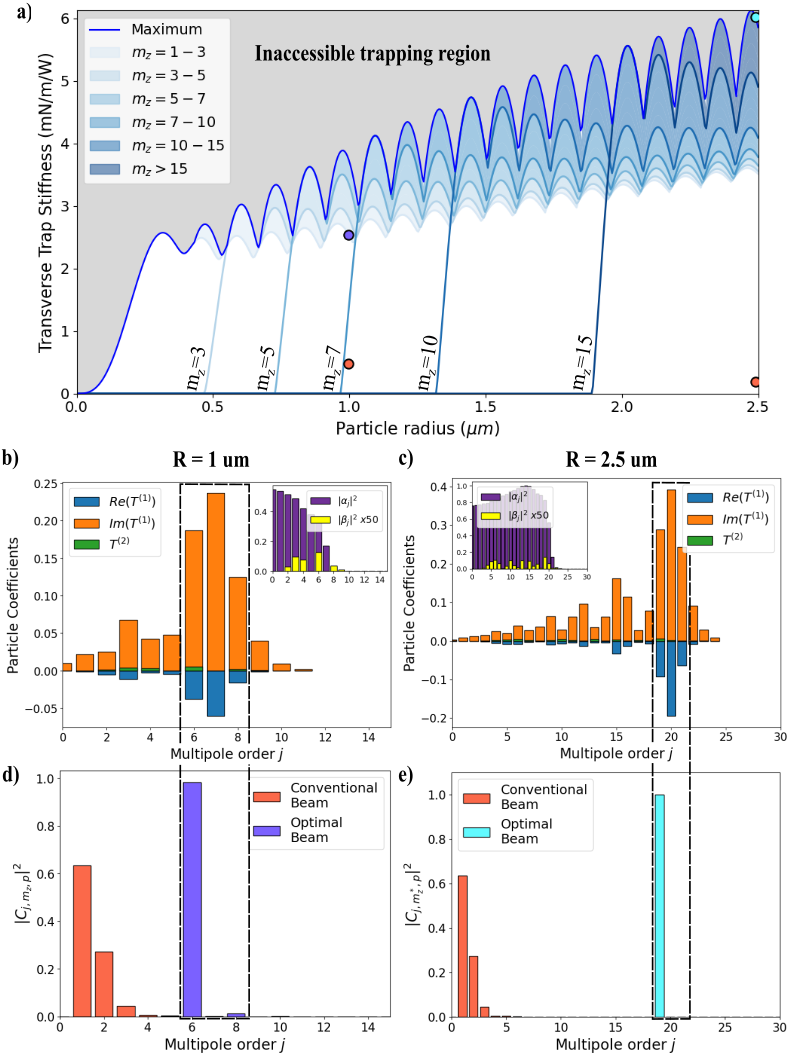}
    \caption{
    Hierarchy of multipolar forces for fields with a given total angular momentum. Panel (a) shows the different regions of transverse trapping stiffness as a function of particle radius and total angular momentum $m_z$. The gray region is inaccessible, while shades of blue indicate areas reachable with increasing $m_z$ (see legend). Solid lines show stiffness obtained with optimal beams for different angular momentum. The white region corresponds to the restriction $m_z=1$. For larger particles the optimal fields can only be achieved with higher angular momenta. This can be explained by observing the multipolar content of the particles: panels (b,c) show Mie force coefficients $T^{(1)}$ (orange/blue) and $T^{(2)}$ (green) for particles of radius $1~\mu$m and $2.5~\mu$m. Insets display the corresponding multipolar scattering coefficients $\alpha_j$ (purple) and $\beta_j$ (x50 yellow). Panels (d,e) compare the multipolar content of optimal beams (purple $m_z=1$, cyan $m_z=19$) with a Gaussian beam (red, $m_z=1$). The dominant multipole follows the maxima of the Mie force coefficients, corresponding to the highest-order multipoles efficiently supported by the sphere. Circles in (a) indicate stiffness values for Gaussian (red) and optimal beams (c,d).
    }
    \label{fig2}
\end{figure}

Before turning to the experimental implementation, we can gain a bit more insight on the optimal beams and their properties by exploring the case of trapping $SiO_2$ spherical particles suspended in water. Fig. \ref{fig2} a) shows the transverse trap stiffness achieved when we restrict the optimal beams to have a given total angular momentum ($m_z$) (blue curves) as a function of the particle radius. We have focused on the transverse trap stiffness, but similar curves can be obtained for trapping along the beam axis. Also, we consider just restoring forces in the radial direction, i.e. forces along the displacement direction ($\textbf{u}=\textbf{v}=\textbf{r}$). Several features can be observed. First, since only trapping on the beam axis is considered in these calculations, beams with larger angular momentum begin to trap only when the particle multipolar content becomes sufficiently large to interact with the multipoles present in the trapping beam.  This behavior has already been reported for conventional LG beams \cite{gomez2024axis,gomez2025optical,o2001axial,sato1991optical}. Furthermore, for big particles, higher angular momenta enables a further increase of the trapping stiffness. For particles with a radius larger than $\sim0.5 \mu$m the maximum trapping limit can only be attained with higher angular momenta.\\

This phenomena can be explained taking a closer look to the force equations. In the case of spherical particles, the force and, therefore, the trapping stiffness matrix, depend on the particle properties through the Mie Force coefficients: $\mathrm{T}^{(1)}_{j}=\alpha_{j}+\alpha_{j+1}^{*}+ 2 ( \alpha_{j} \alpha_{j+1}^{*} + \beta_{j} \beta_{j+1}^{*} )$, and $\mathrm{T}^{(2)}_{j}= | \alpha_{j} |^2 - | \beta_{j} |^2 + \mathrm{Re}(\alpha_{j})$ \cite{bohren2008absorption}. For dielectric particles, $\mathrm{T}^{(2)}_{j}$ is directly related to the scattering cross section. In Fig. \ref{fig2} b) and c) we plot two examples of the Mie Force coefficients for two  $SiO_2$ spheres ($R=1 \mu m$ and $2.5 \mu m$). Together we also plot the scattering coefficients ($\alpha_j$ and $\beta_j$). It is observed that the multipolar scattering coefficients have distinct behaviours, the $\beta_j$ are much smaller than the $\alpha_j$ indicating that most of the light is scattered in the same helicity as the incident beam, and both of them have a sharp cut-off at a given multipolar order ($j_\textrm{cut-off}\sim 10$ for $R=1\mu m$ and $j_\textrm{cut-off}\sim 25$ for $R=2.5\mu m$). On the other hand, the Mie Force coefficients present distinct peaks close to the cut-off multipolar order of the particle.

In Fig. \ref{fig2} d) and e) we present the BSC of a conventional circularly polarized Gaussian beam focused with a microscope objective of NA $= 1.25$. It can be observed that for both sizes of the particles, the Gaussian beam only overlaps with the Mie Force coefficients in a region where the latter are small. Now, in the same figures we also present the elements of the $\mathbf{C}_{m_z,p=1}$ of the maximal eigenvector of the trapping stiffness matrix. In this case, the optimal field adapts to the multipolar content of the sphere. The optimal beam is mainly composed by a single multipole of $j=6$ for the $1\mu m $ particle and at $j=19$ for the $2.5\mu m$ one. Also, while the mode in c) corresponds to a total angular momentum of $m_z=1$, the one in d) corresponds to $m_z=19$. The calculated force increase is about 6-fold for the $1 \mu$m particle, to 33-fold for the $2.5$ one.\\

This analysis also provides an explanation for the behavior of the transverse trap stiffness across different particle sizes, as shown in Fig. \ref{fig1}.b. Given that the multipolar content of the $\mathrm{SiO_2}$ spheres depends on their size—and that higher-order multipoles can only be significantly excited in larger particles—the observed similarity among all beams in the small-particle regime can be attributed to their shared multipolar composition, which is predominantly limited to the first two or three multipoles. As the particle size increases, higher-order multipolar contributions become relevant, and the optimal beam shifts its dominant multipolar component to the next one. This behavior accounts for the discrete jumps observed in the maximum force curve. The periodicity of these jumps is primarily determined by the relation between the wavelength of the trapping laser and the size of the particle \cite{stilgoe2008effect}.\\

This multipolar analysis appears as a crucial tool to explain the difference between, not only the enhanced trapping of our optimal beams, but also to account for the trap stiffness enhancement of high order LG beams with respect to conventional Gaussian beams for large particles \cite{gomez2024axis,o2001axial,sato1991optical}.

\section{Experimental realization of the optimal beams}\label{sec3}
\begin{figure}
    \centering
    \includegraphics[width= 0.9\textwidth]{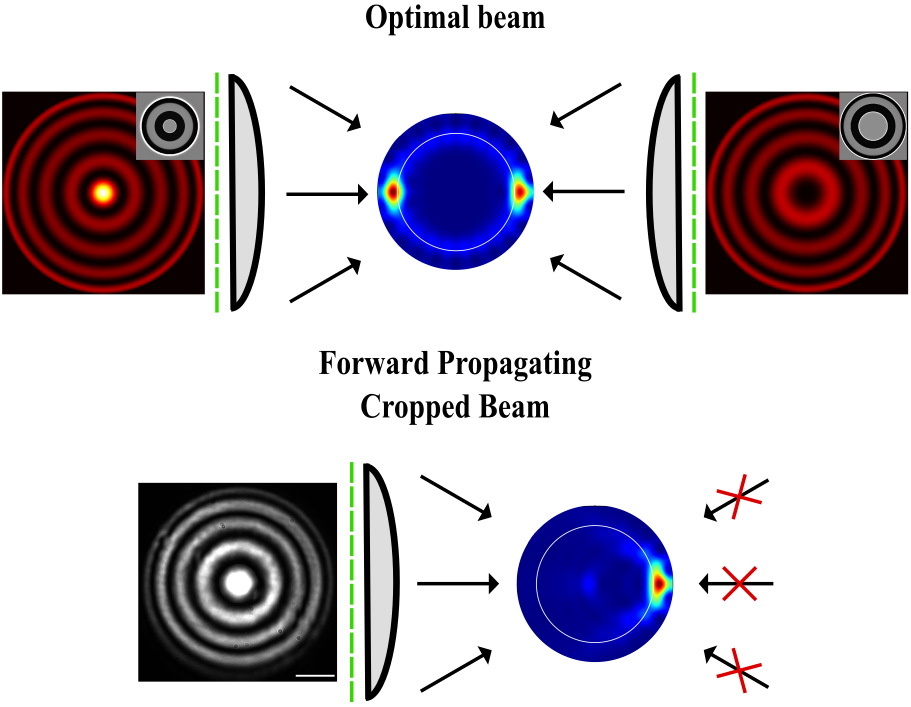}
    \caption{Realization of the Forward Propagating Cropped (FPC) beams. Schematic representation of how the FPC beams are created. At the top the optimal beams require the combination of both forward and backward propagating fields which experimentally is very challenging. While, at the bottom, the approximation composed solely by forward propagating fields.
    }
    \label{fig3}
\end{figure}

This approach is not only a theoretical tool, but, as mentioned earlier, has an operational meaning for the design of experiments. To show this, we conducted experiments using these techniques on the trapping of silica  ($SiO_2$) particles in water. The key to understand the experimental procedure is realizing that the optimal beams are realized with non-singular multipolar fields, i.e. they must have a finite value at the origin. Then, it is always possible to fully decompose these fields as plane waves \cite{wittmann2002spherical}. Even though multipolar fields of well-defined helicity have a preferred propagation direction \cite{molezuelas2024characterizing}, in order to maximize the multipolar content, we should use counter-propagating beams, with both forward and backward propagating fields (Figure \ref{fig3} optimal beam). On the other hand, most experiments in optical trapping use a single beam configuration, given that losses, misalignment and aberrations would may have a negative impact on the trapping. Then, we designed a single beam experimental realization that approximates the multipolar content of the optimal beam while using only forward-propagating plane waves. Figure \ref{fig3} visually shows the realization of the Forward-Propagating Cropped (FPC) beams. One can observe on the top row the optimal beams that are composed by both forward and back propagating fields. Our single beam realization simply crops the backward propagating plane waves (bottom row). Although this FPC beam does not achieve the maximum possible trap stiffness predicted by our theory, it enhances the trap stiffness of conventional beams by up to an order of magnitude (Fig. \ref{fig1}.b).\\

To generate the forward-propagating cropped beam, we used a phase spatial light modulator (SLM) operating in reflection mode to modulate both the phase and amplitude of an incoming Gaussian beam. In order to impart an angular momentum of $m_z$ to the beam, we must add an azimuthal phase that varies by $2 \pi \ell$ in the SLM, where $\ell$ is the topological charge or also called Orbital Angular Momentum (OAM). The resulting angular momentum is $m_z=\ell+p$, depending on the helicity $p$ of the field. The modulated field was then projected through a 4f lens system onto the back focal aperture of the trapping objective, allowing us to directly control both the phase and amplitude of the beam and, consequently, the multipolar content of the trapping beam. This was achieved using multipolar theory and the aplanatic lens model of an objective \cite{zambrana2012excitationsinglemult,zambrana2015control}. An example of the experimental intensity at the back-focal aperture of the objective used to create the FPC beams in this work is shown in Fig. \ref{fig3}.

The light scattered by the particle was then collected by another objective and projected onto a four-quadrant photodetector, where a power spectral density (PSD) analysis was performed. From a Lorentzian fit, we extracted the trap stiffness \cite{gittes1997signals,ghislain1994measurement,sarshar2014comparative,resnick2017optical}.

We trapped silica particles of $2\mu m$ and $3.17\mu m$ in diameter, suspended in water, using a $976 nm$ laser focused by an oil-immersion objective with NA$= 1.25$. We measured the trap stiffness for different powers of the trapping laser at the back focal aperture of the objective and for different total angular momentum ($m_z$) of both the conventional beams and our Forward-Propagating Cropped beams.

\begin{figure}
    \centering
    \includegraphics[width=\textwidth]{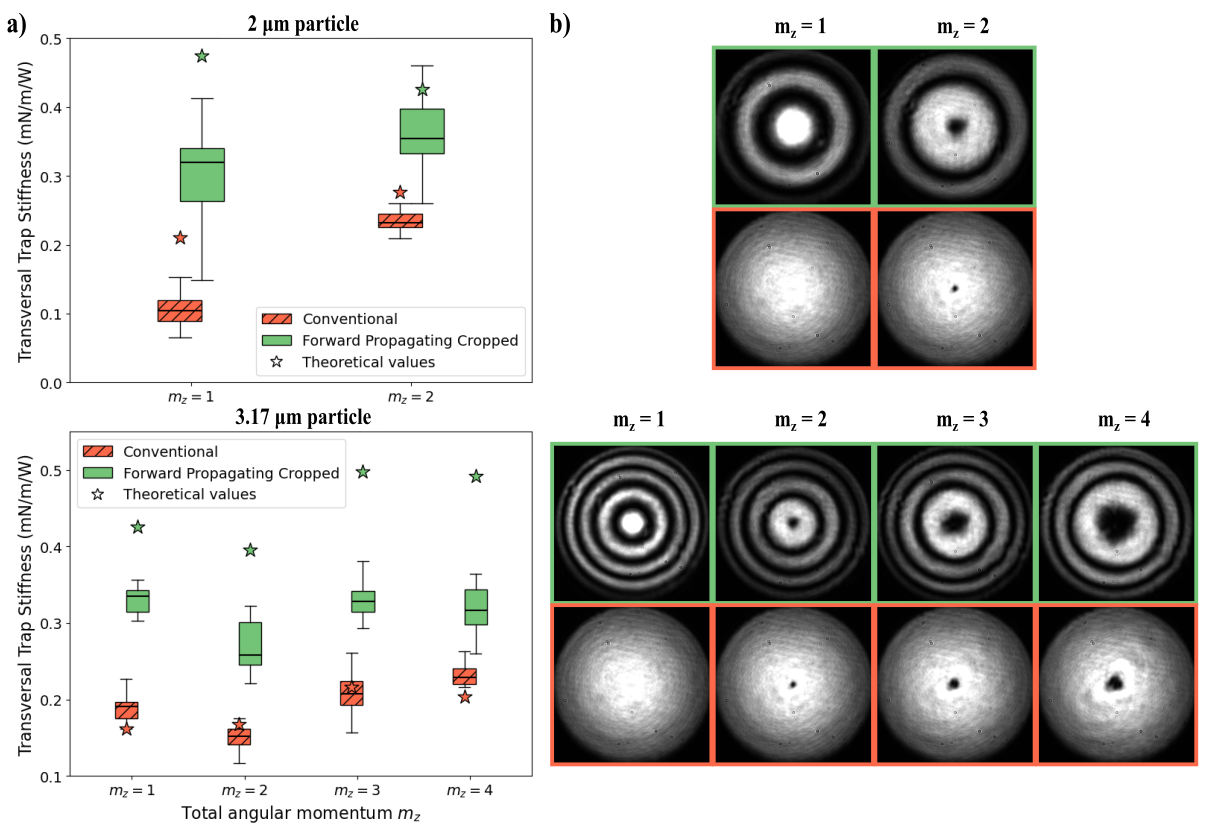}
    \caption{Experimental enhancement of the Forward Propagating Cropped (FPC) beams. (a) Box plot of the transversal trap stiffness per unit of power obtained experimentally for particles of 2 and 3.17 $\mu m$ of diameter for different topological charges $\ell$ of the trapping beam. A general enhancement of the trap stiffness is observed for the FPC beams (green). (b) Experimental intensity distribution at the back focal aperture used to generate the different conventional (orange) and FPC (green) beams for the different topological charges. A ring-like structure that depends on the size of the particle is observed for all the FPC beams while the conventional vortex beams remain the same regarding of the size of the particle. The box plots represent the data distribution, with the box extending from the first quartile (Q1) to the third quartile (Q3), the central line indicating the median, and the whiskers extending to the most extreme data points within 1.5 times the interquartile range (IQR).}
    \label{fig4}
\end{figure}

Figure \ref{fig4} shows the experimental data recorded for the experimental FPC beams together with the conventional beams. Figure \ref{fig4}.a shows a box plot of the transverse trap stiffness per unit power, obtained from 15–20 (depending on the trapping beam) different values of the beam power at the back-focal aperture of the objective. The data were recorded for particles of 2 $\mu m$ and 3.17 $\mu m$ of diameter with conventional beams (orange) and our FPC beams (green) with different values of the angular momentum ($m_z$). Each of the stars of the corresponding color marks the theoretical values predicted for each beam, taking into account the total efficiency measured of the microscope objective (60\% on average for the measured fields). For every data set, the FPC beams outperform the conventional ones with enhancements of 3.1 and 1.6 for the 2 $\mu m$ particles and enhancements of 2.2, 1.6, 1.8 and 1.5 for the 3.17 $\mu m$ particles for the different angular momentum, respectively. Figure \ref{fig4}.b presents the experimental intensity distribution at the back focal aperture used to generate the different conventional (orange) and FPC (green) beams for the different angular momentum.
We can also observe a clear difference between our FPC beams and conventional vortex beams: the FPC beams display a similar ring-like distribution, with the number and size of rings depending on the size of the trapped particle, while the conventional beams remain the same regardless of the particle.

These experimental results are in very good agreement with the theoretical predictions, showcasing that, even though the experimental approximation of the FPC beams does not achieve the enhancement theoretically predicted for the optimal beams, they are still a experimentally feasible way to obtain an increase in trap stiffness without the need for iterative live optimization \cite{butaite2024photon,taylor2015enhanced,taylor2017optimizing}.

Our theoretical framework predicts that increasing the angular momentum should further enhance the trap stiffness, particularly for particles around 3.17 $\mu$m in diameter. While this trend is not fully realized experimentally for higher values of $m_z$, the observed deviations provide valuable insight into the role of optical aberrations in structured trapping fields. In particular, spherical aberrations in the microscope objective, which are more pronounced for light propagating through the outer regions of high-NA lenses \cite{butaite2024photon}, are expected to disproportionately affect beams with larger angular momentum. Additional contributions may arise from aberrations at the water–glass interface \cite{da2024tailoring} and from residual imperfections in the control of the multipolar content of the forward-propagating cropped beams, which, as highlighted in this work, critically determines the achievable trap stiffness. Therefore, these results not only validate the predicted trends but also identify clear pathways for further optimization of structured optical trapping systems.

\section{Conclusions}
In this work, we establish an analytical framework that determines the fundamental upper bound to optical forces with the knowledge of the scattering matrix of the trapped object. By expressing the optical force in a helicity-resolved multipolar basis, we show that small-displacement forces—and thus trap stiffness—admit a closed analytical representation in which the properties of the optical field are fully separated from those of the trapped object. Within this formulation, the maximum eigenvalue of the resulting stiffness matrix defines an absolute upper bound on the achievable trapping strength, while the associated eigenvector uniquely identifies the optimal field configuration required to attain it. We have shown a direct application of this method to the forces exerted by cylindrically symmetric beams on homogeneous spherical particles, a common situation in many optical trapping experiments. Extensions to non-cylindrically symmetric beams, such as linearly polarized focused beams can be directly implemented, just by expanding the trapping stiffness and force matrices. 

However, the work presented here provides valuable insight on the role of angular momentum of light on optimizing optical forces: Applying this framework to spherical SiO$_2$ particles in water reveals that optimal trapping fields are dominated by a single multipolar component occurring near the cutoff order set by the particle size. This multipolar perspective provides a unified interpretation of stiffness enhancement mechanisms, including the improved performance of high-order Laguerre–Gaussian beams relative to conventional Gaussian beams observed in previous studies \cite{o2001axial,gomez2024axis,sato1991optical}. Also, for larger and larger particles the maximum trapping stiffness achievable is reached through a hierarchy of angular momentum beams.

From a practical standpoint, we demonstrate that experimentally accessible forward-propagating approximations capture a large fraction of the theoretically predicted enhancement. Our measurements confirm substantial increases in trap stiffness—up to a factor of three—without relying on iterative optimization, thus validating the predictive power of the analytical framework introduced here.

The remaining difference between theoretical limits and the present experimental realizations reflects practical constraints such as finite numerical aperture, residual aberrations, and incomplete control of the multipolar content. Addressing these factors provides clear routes towards approaching the fundamental limits identified in this work.

More broadly, the methodology developed here is general and can be extended to complex particle geometries, heterogeneous environments, and non-cylindrically symmetric optical fields through the use of the scattering matrix. By enabling systematic identification of optimal field configurations from first principles, this approach establishes a route toward the rational design of light–matter interactions, with potential applications ranging from levitated optomechanics to precision manipulation of biological systems with reduced optical damage.

\section{\label{sec:Acknowled}Acknowledgments}
E.A.G., Q.P.S., J.T.F., A.M.M., S.J.R., M.V. and G.M.T. acknowledge support from the CSIC Research Platform on Quantum Technologies, from IKUR Strategy under the collaboration agreement between Ikerbasque Foundation and DIPC/MPC on behalf of the Department of Education of the Basque Government. A.M.M. also acknowledges the financial support from the Department of Education of the Basque Government through the Non-Doctoral Researcher Pre-Doctoral Formation Program, Grant No. PRE 2024 2 0180.

\bibliography{bibliography}

\end{document}